\documentclass[aps,pre,floats,twocolumn,showpacs,superscriptaddress]{revtex4}

\usepackage{graphicx}
\usepackage{dcolumn}
\usepackage{bm}
\usepackage{booktabs}
\usepackage{color,times}

\begin{document}
\newcommand{\tbox}[1]{\mbox{\tiny #1}}
\newcommand{\bra}{\left\langle}
\newcommand{\ket}{\right\rangle}

\title{Weighted random--geometric and random--rectangular graphs: 
Spectral and eigenfunction properties of the adjacency matrix}

\author{L. Alonso}
\email{lazarus.alon@gmail.com}
\affiliation{
Instituto de F\'isica, Benem\'erita Universidad Aut\'onoma de Puebla, Apartado Postal J-48, Puebla 72570, Mexico
}
\author{J. A. M\'endez-Berm\'udez}
\email{jmendezb@ifuap.buap.mx}
\affiliation{
Instituto de F\'isica, Benem\'erita Universidad Aut\'onoma de Puebla, Apartado Postal J-48, Puebla 72570, Mexico
}
\author{A. Gonz\'alez-Mel\'endrez}
\affiliation{
Instituto Tecnol\'ogico de Estudios Superiores de Occidente, Tlaquepaque 45604, Jalisco, Mexico
}
\author{Yamir Moreno}
\affiliation{Instituto de Biocomputaci\'on y F\'{\i}sica de Sistemas Complejos (BIFI), Universidad de Zaragoza, 50018 Zaragoza, Spain}
\affiliation{Departamento de F\'{\i}sica Te\'orica. University of Zaragoza, Zaragoza E-50009, Spain}
\affiliation{Institute for Scientific Interchange, ISI Foundation, Turin, Italy} 

\date{\today}

\begin{abstract}
Within a random-matrix-theory approach, we use the nearest-neighbor energy level spacing 
distribution $P(s)$ and the entropic eigenfunction localization length $\ell$ to study spectral and 
eigenfunction properties (of adjacency matrices) of weighted random--geometric and 
random--rectangular graphs.
A random--geometric graph (RGG) considers a set of vertices uniformly and independently 
distributed on the unit square, while for a random--rectangular graph (RRG) the embedding 
geometry is a rectangle. 
The RRG model depends on three parameters: The rectangle side lengths $a$ and $1/a$, 
the connection radius $r$, and the number of vertices $N$. 
We then study in detail the case $a=1$ which corresponds to weighted RGGs and explore 
weighted RRGs characterized by $a\sim 1$, i.e.~two-dimensional geometries, but also approach 
the limit of quasi-one-dimensional wires when $a\gg1$. 
In general we look for the scaling properties of $P(s)$ and $\ell$ as a function of $a$, $r$ and $N$. We find that the ratio $r/N^\gamma$, with $\gamma(a)\approx -1/2$, fixes the properties 
of both RGGs and RRGs. Moreover, when $a\ge 10$ we show that spectral and eigenfunction 
properties of weighted RRGs are universal for the fixed ratio $r/{\cal C}N^\gamma$, with
${\cal C}\approx a$.
\end{abstract}

\maketitle

\section{Introduction and graph model}

Complex networked systems are made up by many units (be them individuals, atoms, genes, plants, etc) that interact among them following often non-trivial patterns. These patterns can be mapped into complex networks \cite{review2006}. The study of the interaction topology has provided deep insights into the dynamics and functioning of complex systems found in biological, social, technological and natural domains. In general, networks are not embedded in any physical or geographical space, but often, we need to preserve the spatial component \cite{B11} of the interaction networks. In such cases, a widely used class of models is the Random Geometric Graph  (RGGs) \cite{DC02, P03}. These graphs share many properties with Random Graphs  \cite{G59}, like a homogeneous degree distribution, but contrary to the latter, they might also show high clustering and a larger mean average path length. Additionally, a time dependent network topology is obtained from RGGs by allowing nodes to move in and out of each other's contact range according to a random walk mobility model. These properties have made RGGs very useful to study the structure and dynamics of spatially embedded complex systems, with applications as diverse as the characterization of the spatial connectivity when the population density is non-uniform \cite{WG09}, synchronization phenomena \cite{DGMN09}, wireless ad-hoc communications \cite{Nekovee07} and disease dynamics \cite{TG07, BAK08}.

\begin{figure*}[th]
\centering
 \includegraphics[scale = .5]{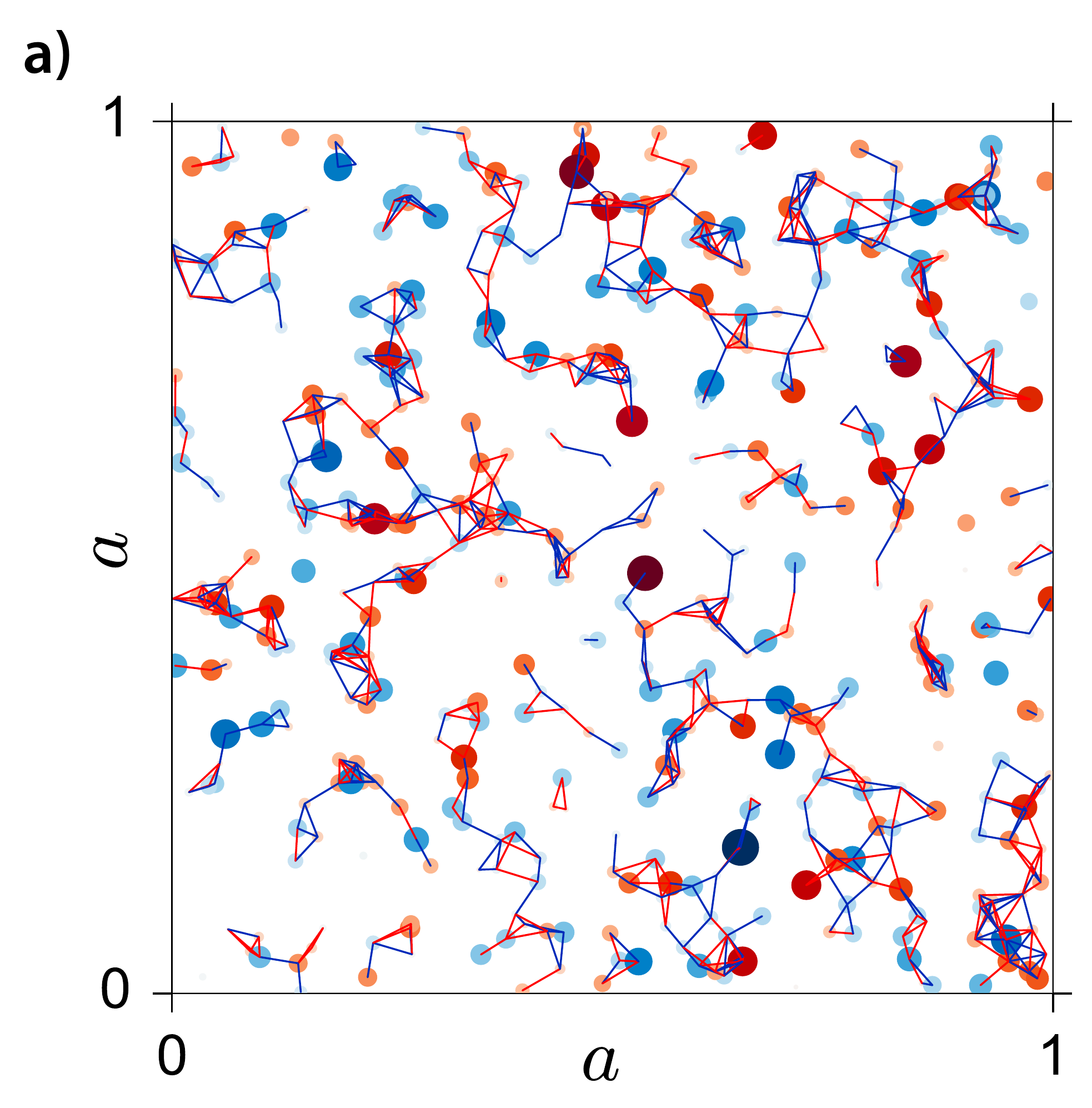}
 \includegraphics[scale = .5]{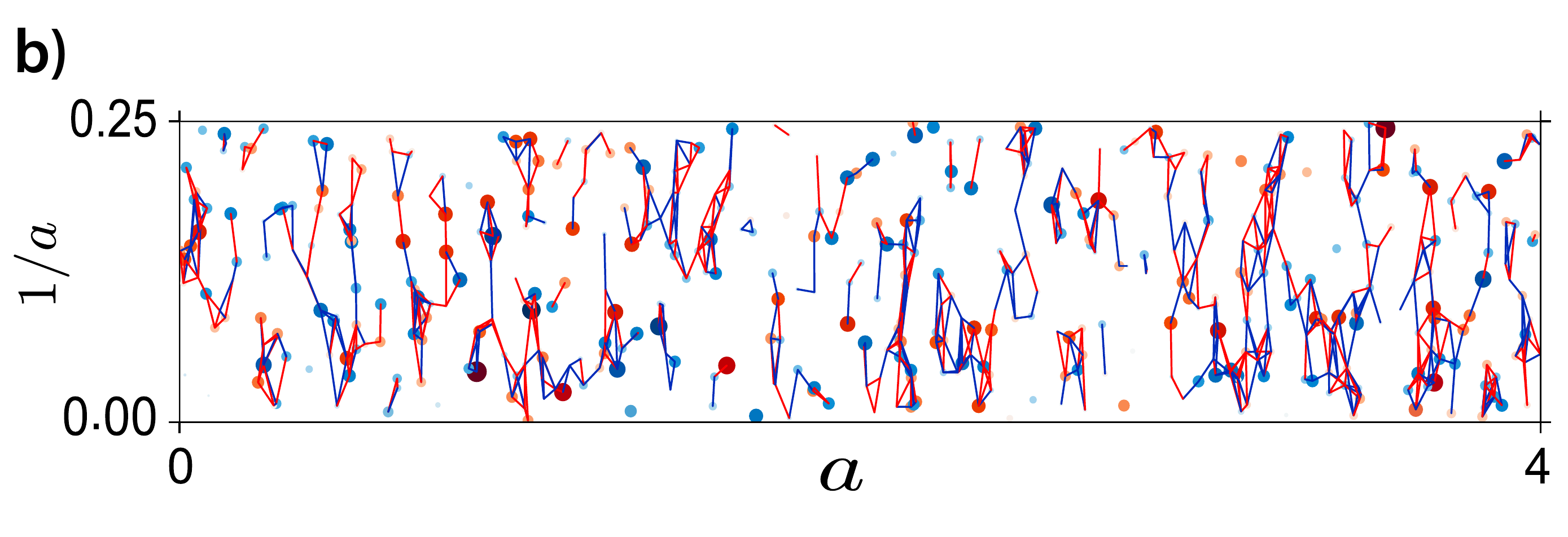}
\caption{(Color online) (a) A weighted RGG with $N=500$ and $r=0.05$ (equivalently a weighted 
RRG with $a=1$). (b) A weighted RRG with $N=500$, $r=0.05$, and  $a=4$. Here, since vertices and edges have weights given by Gaussian random variables with zero mean, we draw vertices and edges in red (blue) if the weight is negative (positive). Also, the larger the size (width) of the vertex (edge) the larger its weight magnitude.}
\label{Fig1}
\end{figure*}

RGGs are however defined in a two-dimensional plane. In some situations, like when modeling disease transmission between plants, or some cities, it is convenient to have the freedom to tune the shape of the space in which nodes are embedded. Random rectangular graphs (RRGs) were recently introduced as a generalization of RGGs~\cite{ES15} to allow for the previous flexibility in the geographical space. A RRG is defined as a set of $N$ vertices (nodes) uniformly and independently 
distributed on a rectangle of unit area. Here, two vertices are connected by an edge if their Euclidian 
distance is $\leq r$. Therefore, this random graph model depends on three parameters: The rectangle 
side lengths $a$ and $1/a$, the connection radius $r$, and the number of vertices $N$. 
Note that when $a\sim 1$ the vertices are placed in a two-dimensional box, while if $a\gg1$
they are embedded in a quasi-one-dimensional geometry. 
When $a=1$, i.e.~when the rectangle becomes the unit square, the RGG model is recovered. 
As examples, in Fig.~\ref{Fig1} we show a RGG (upper panel) and 
a RRG with $a=4$ (lower panel), both having $N=500$ and $r=0.05$.

Regardless of the very recent introduction of RRGs~\cite{ES15} there are already several studies 
available on them~\cite{ES15,EC15,ES16,EMSM16}. Topological properties such as the average degree, the degree distribution, the average path length, and the clustering coefficient were reported in~\cite{ES15}; while dynamical properties of processes occuring on top of these graphs were studied in~\cite{EC15} (synchronizability),~\cite{ES16} (consensus dynamics), and~\cite{EMSM16}
(disease spreading). The main conclusion from these works is that the properties of RRGs and the dynamics of the system strongly depend on the rectangle elongation and thus, they can be drastically different to those of RGGs. For example, for $a\gg 1$ the
RRG becomes `large-world'; i.e.~its connectivity decays to zero. 

However, to the best of our knowledge, the study of spectral and eigenfunction properties of RRGs 
is still lacking (although, some spectral properties of RGGs have already been reported, both 
theoretically~\cite{BEJ07} and numerically~\cite{BEJ07,NGB14,DGK16}). Spectral properties are important in the study of several dynamical processes on networks, as in many cases, they determine the critical properties of the systems under study. Therefore, in this paper we undertake this task and perform a systematic analysis of 
the spectral and eigenfunction properties of RRGs within a random-matrix-theory (RMT) approach. 
Specifically, we use the nearest-neighbor energy level spacing distribution $P(s)$ and the entropic
eigenfunction localization length $\ell$ to characterize spectral and eigenfunction properties of the 
adjacency matrices of weighted RRGs. In doing so, we first analyze in detail weighted RGGs.
In general, we look for the scaling properties of $P(s)$ and $\ell$ as a function of $a$, $r$ and $N$.
We report that the ratio $r/N^\gamma$, with 
$\gamma(a)\approx -1/2$, fixes the properties of both RGGs and RRGs.
In addition, we also demonstrate that for $a\ge 10$ the spectral and eigenfunction 
properties of RRGs are universal for the fixed ratio $r/{\cal C}N^\gamma$, with
${\cal C}\approx a$.

Here, furthermore, we consider an important addition to the standard~\cite{ES15,EC15,ES16,EMSM16}
RRG model: Weights for vertices and edges. In fact, the examples shown in Fig.~\ref{Fig1} correspond 
to weighted random graphs (see the caption for the color coding).
Our motivation to include weights to the standard RRG model is three-fold:
First, we would like to consider more realistic graphs in which vertices and edges are not equivalent;
therefore their corresponding adjacency matrices are not just binary, see e.g.~\cite{BBPV04}.
Second, we want to retrieve well known random matrices in the appropriate limits in order to use
RMT results as a reference. And third, we want to effectively approach the limit of $r\to 0$, i.e., the limit of strongly sparse graphs,
which is not accessible with binary adjacency matrices since binary matrices become the null matrix 
in that limit. Specifically, for our weighted RRG model, the non-vanishing elements of the corresponding adjacency 
matrices are statistically independent random variables drawn from a normal distribution with zero 
mean $\bra A_{ij} \ket=0$ and variance $\bra |A_{ij}|^2 \ket=(1+\delta_{ij})/2$. Accordingly, a diagonal adjacency random matrix is obtained for $r=0$ (known in RMT as the Poisson 
case), whereas the Gaussian Orthogonal Ensemble (GOE) is recovered when $r=a$.

A study of spectral properties of RGGs, also within a RMT approach, has 
been recently reported in Ref.~\cite{DGK16}. However, in contrast to our study, the authors of \cite{DGK16}
analyzed binary adjacency matrices and focused on parameter combinations for which the graph is
in the weak sparse regime only. Below we use exact numerical diagonalization to obtain the eigenvalues $E^m$ and eigenfunctions
$\Psi^m$ ($m=1\ldots N$) of the adjacency matrices of large ensembles of weighted random graphs 
characterized by $a$, $r$, and $N$.

\section{Weighted random geometric graphs}

In order to set up our approach with a somewhat simpler (in the sense that it contains one parameter less), but extensively studied, model we first study weighted RGGs.

\subsection{Spectral properties}

\begin{figure}[h]
\centering
 \includegraphics[scale = .4]{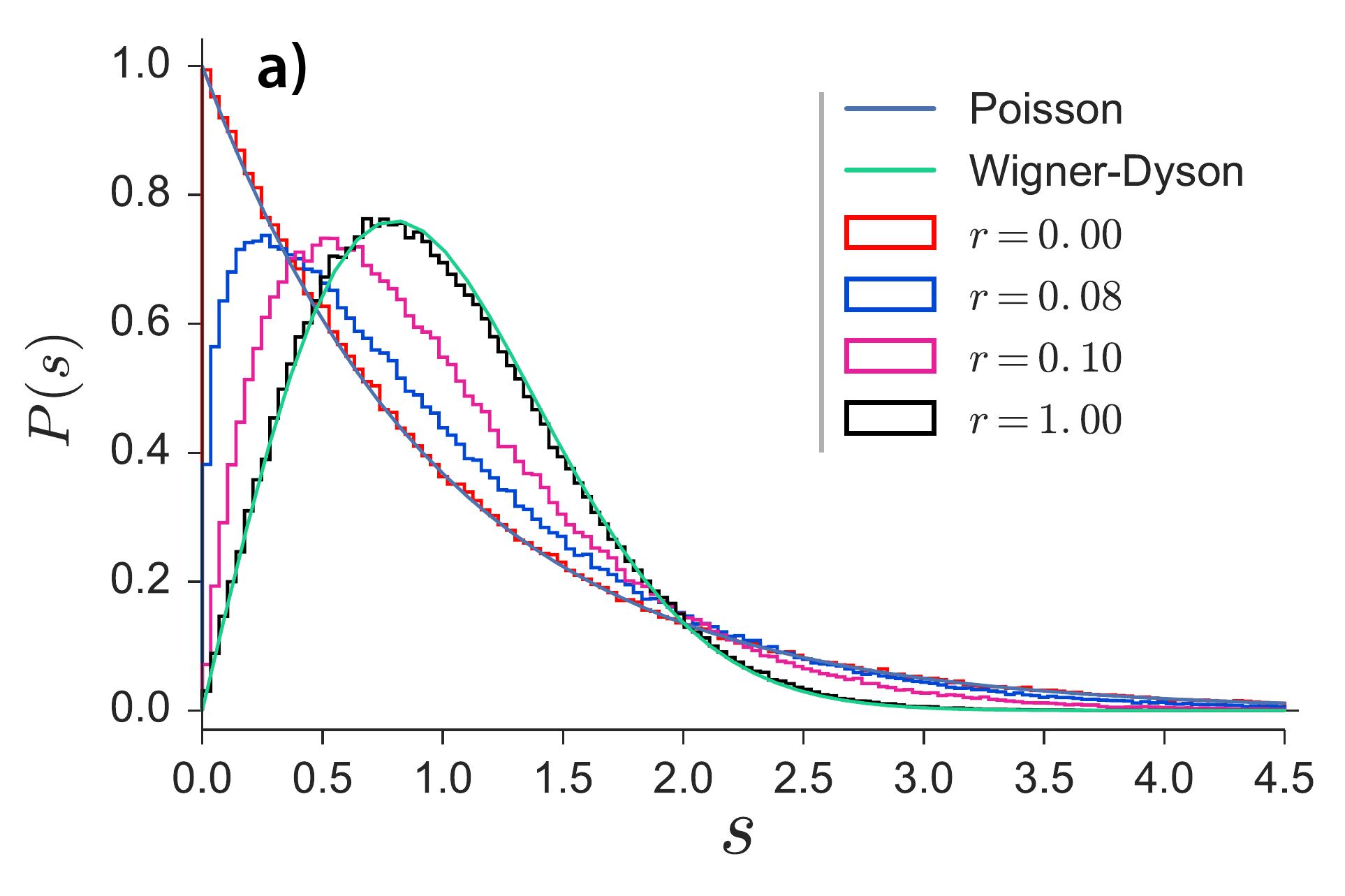}
 \includegraphics[scale = .4]{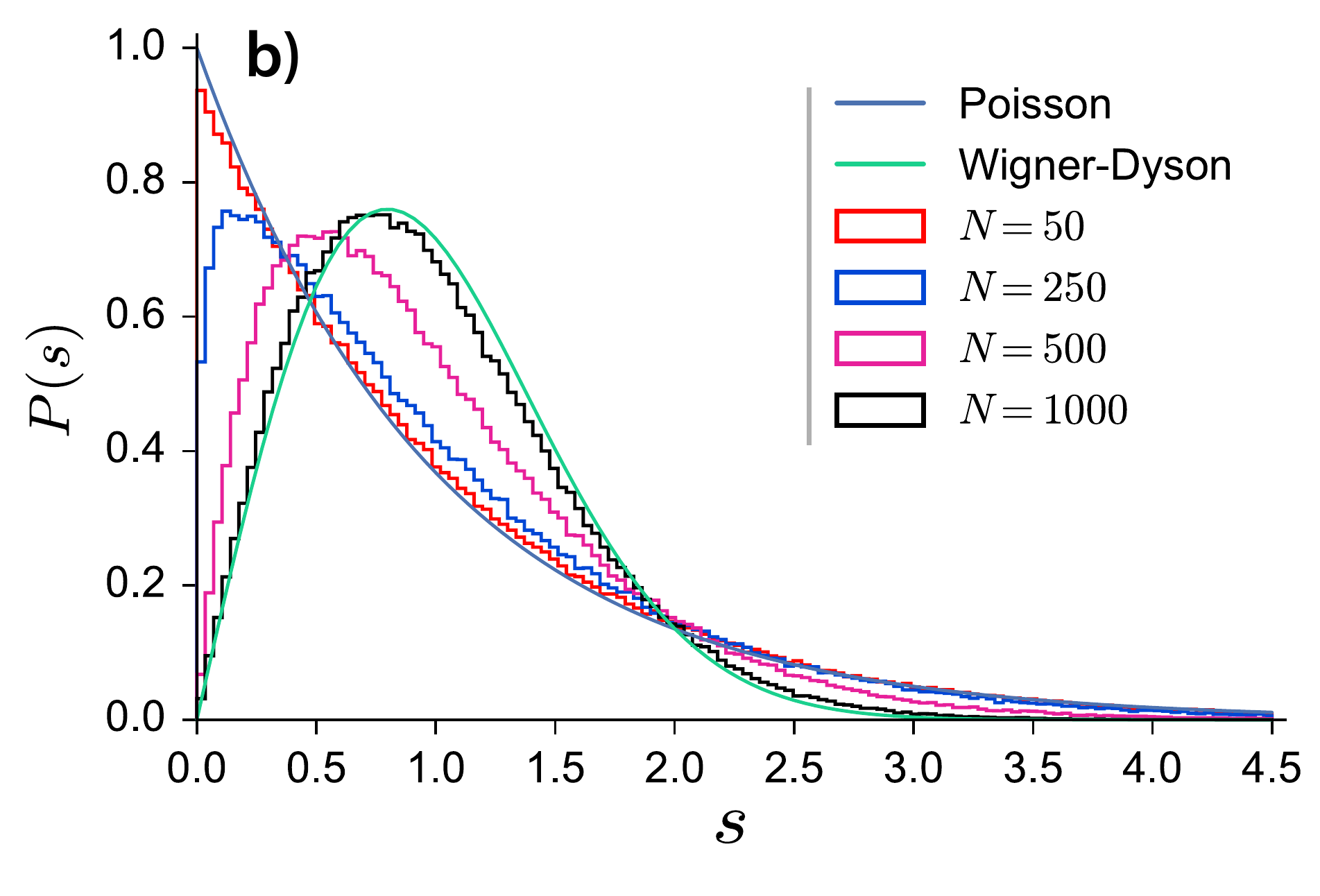}
\caption{(Color online) Nearest-neighbor energy level spacing distribution $P(s)$ for weighted RGGs.
In (a) $N = 500$ and $r=0$, 0.08, 0.1, and 1. In (b) $r = 0.1$ and $N=50$, 250, 500, and 1000.
Full lines correspond to Poisson and Wigner-Dyson distribution functions given by Eqs.~(\ref{P}) 
and (\ref{WD}), respectively.}
\label{Fig2}
\end{figure}

As anticipated, here we use the nearest-neighbor energy level spacing distribution 
$P(s)$~\cite{metha} to characterize the spectra of weighted RGGs. For $r=0$, i.e., when the vertices 
in the weighted RGG are isolated, the corresponding adjacency matrices are diagonal and, regardless of the size of the graph, $P(s)$ follows the exponential distribution,
\begin{equation}
\label{P}
P(s) = \exp(-s) \ ,
\end{equation}
which is better known in RMT as Poisson distribution. In the opposite limit, $r=1$, when the weighted graphs 
are fully connected, the adjacency matrices become members of the GOE (full real symmetric random
matrices) and $P(s)$ closely follows the Wigner-Dyson distribution,
\begin{equation}
\label{WD}
P(s) = \frac{\pi}{2} s \exp \left(- \frac{\pi}{4} s^2 \right) \ .
\end{equation}
Thus, for a fixed graph size, by increasing $r$ from zero to one, the shape of $P(s)$ is 
expected to evolve from the Poisson distribution to the Wigner-Dyson distribution. Moreover, due to 
the increase in the density of nodes, a similar transition is expected to occur by increasing 
$N$ for a fixed connection radius. In Fig.~\ref{Fig2} we explore both scenarios.

We construct histograms of $P(s)$ by using $N/2$ unfolded spacings \cite{metha},
$s_m=(E^{m+1}-E^m)/\Delta$, around the band center of a large number of graph realizations
(such that all histograms are constructed with $5\times 10^5$ spacings). 
Here, $\Delta$ is the mean level spacing computed for each adjacency matrix.
Figure~\ref{Fig2} presents histograms of $P(s)$ for the adjacency matrices of weighted RGGs:
In Fig.~\ref{Fig2}(a) the graph size is fixed to $N = 500$ and $r$ takes the values 0, 0.08, 0.1, and 1.
In Fig.~\ref{Fig2}(b) the connection radius is set to $r = 0.1$ while $N$ increases from 50 to 1000.
In both figures, as anticipated, one can see that $P(s)$ evolves from Poisson to Wigner-Dyson 
distribution functions (also shown as reference). Transitions from Poisson to Wigner-Dyson in the spectral statistics have also been reported for adjacency matrices corresponding to other complex network models in 
Refs.~\cite{EE92,JMR01,ZX00,GGS05,SKHB05,JKBH08,BJ07,JB08,ZYYL08,J09,MAM15}.

Now, in order to characterize the shape of $P(s)$ for weighted RGGs we use the Brody distribution
\cite{B73,B81} 
\begin{equation}
\label{B}
P(s) = (\beta +1) a_{\beta} s^{\beta}  \exp\left(- a_{\beta} s^{\beta+1}\right) \ ,
\end{equation}
where $a_{\beta} = \Gamma[(\beta+2)/(\beta+1)]^{\beta+1}$, $\Gamma(\cdot)$ is the gamma 
function, and $\beta$, known as Brody parameter, takes values in the range $[0,1]$.
Equation~(\ref{B}) was originally derived to provide an interpolation expression for $P(s)$ in the 
transition from Poisson to Wigner-Dyson distributions. In fact, $\beta=0$ and $\beta=1$ in Eq.~(\ref{B})
produce Eqs.~(\ref{P}) and (\ref{WD}), respectively. We want to remark that, even though the
Brody distribution has been obtained through a purely phenomenological approach and the Brody
parameter has no decisive physical meaning, it serves as a measure for the
degree of mixing between Poisson and GOE statistics.
In particular, as we show below, the Brody parameter will allows us to identify the onset of the 
delocalization transition and the onset of the GOE limit in RGGs.
It is also relevant to mention that the Brody distribution has been applied to study other complex 
networks models~\cite{BJ07,JB08,ZYYL08,J09,JB07,MAM15,DGK16}.

\begin{figure}[h]
\centering
 \includegraphics[scale = .4]{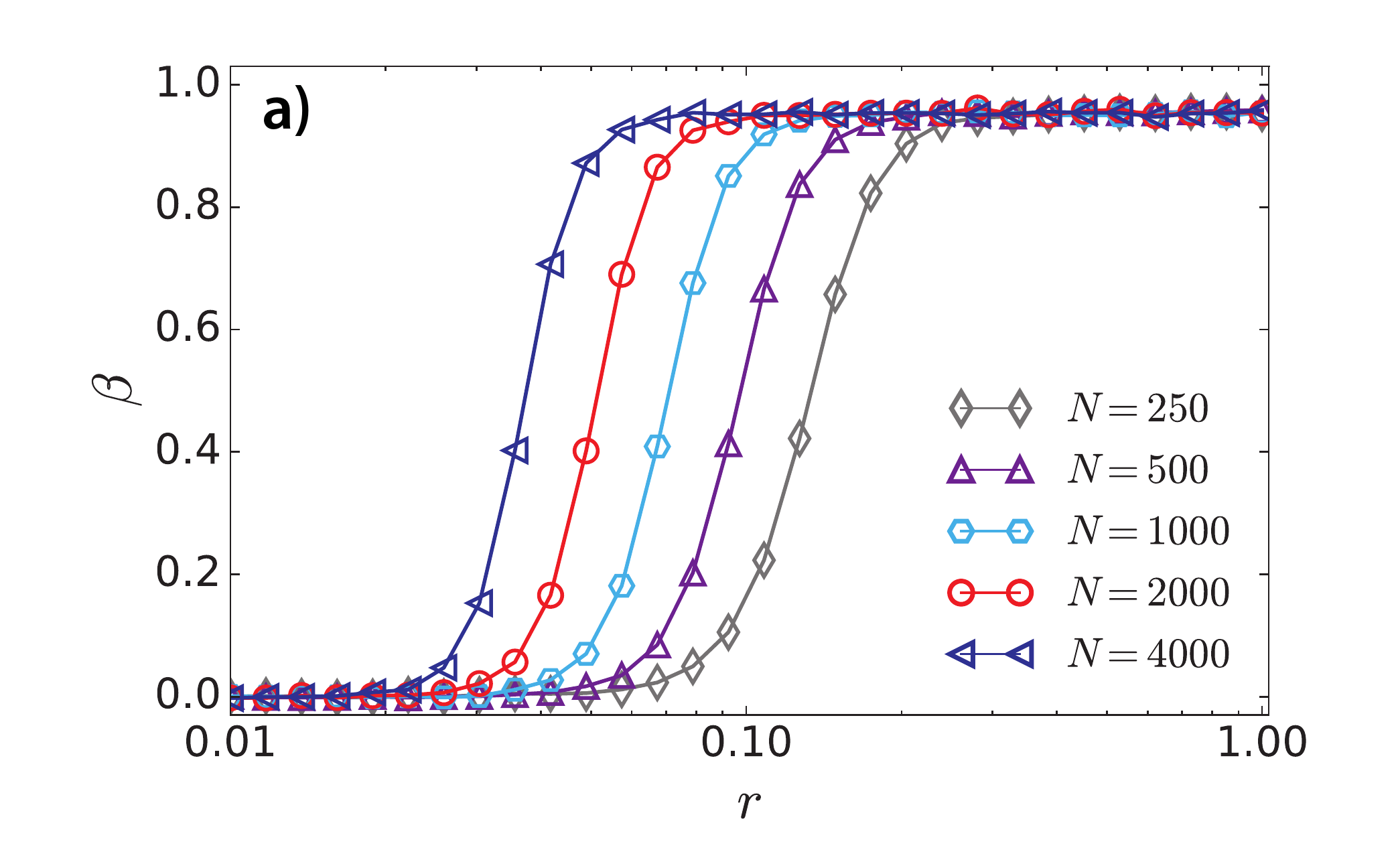}
 \includegraphics[scale = .4]{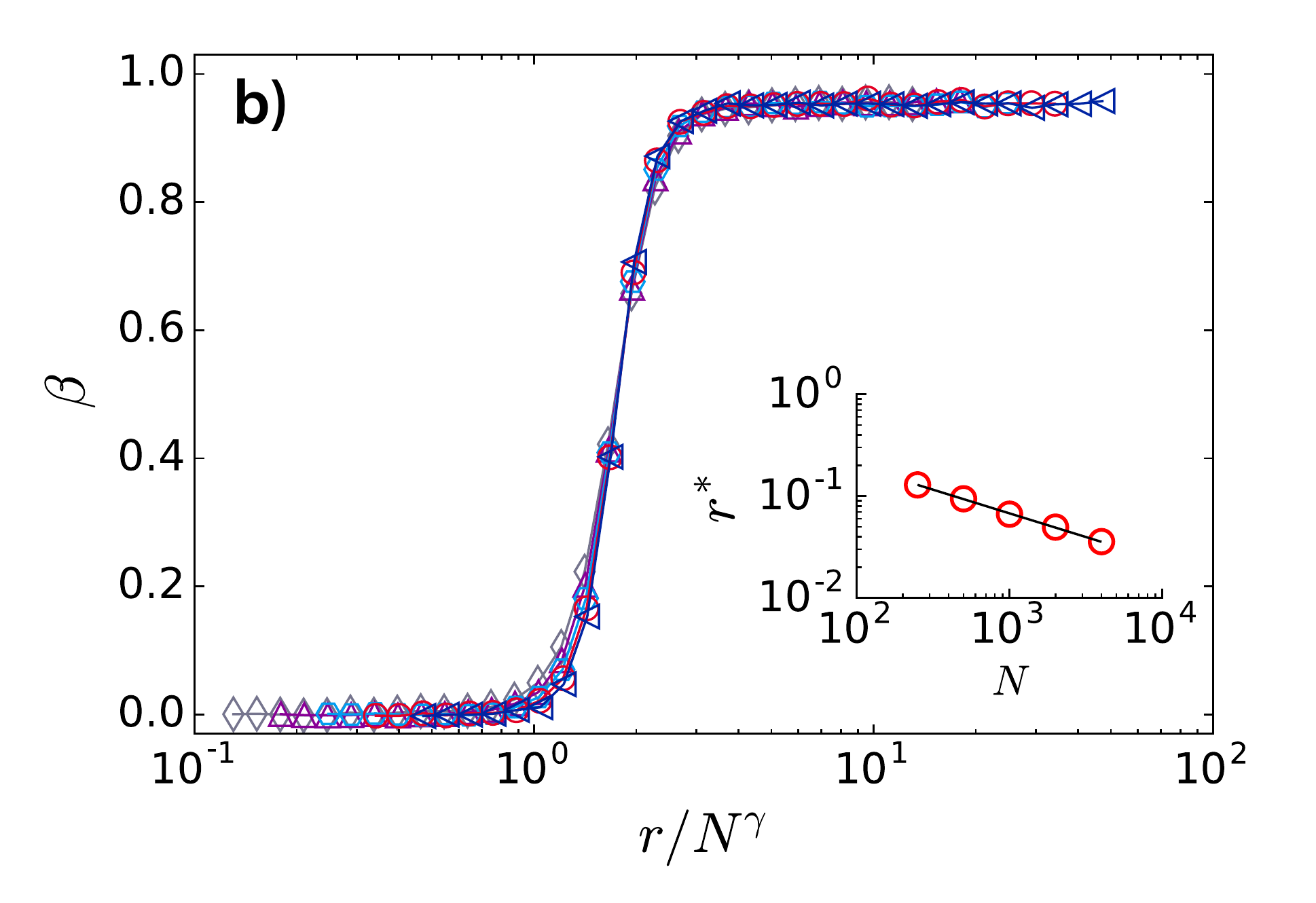}
\caption{Brody parameter $\beta$ as a function of (a) $r$ and (b) $r/N^\gamma$ for weighted 
RGGs of sizes ranging from $N=250$ to 4000. Inset in (b) $r^*$ vs.~$N$. Full black line is the 
fit of the data with Eq.~(\ref{scalingEq}) giving $\mathcal{C}=1.678\pm 0.067$ and 
$\gamma=-0.465\pm 0.006$.}
\label{Fig3}
\end{figure}

We now perform a systematic study of the Brody parameter as a function of the graph parameters
$r$ and $N$. To this end, we construct histograms of $P(s)$ for a large number of 
parameter combinations to extract systematically the corresponding values of $\beta$
by fitting them using Eq.~(\ref{B}) (not shown here). Figure~\ref{Fig3}(a) reports $\beta$ versus $r$ for five different graph sizes. Notice that in all cases the behavior of $\beta$ is similar: $\beta$ shows a smooth transition from zero (Poisson 
regime) to one (Wigner-Dyson or GOE regime) when $r$ increases from $r\ll 1$ (mostly isolated
vertices) to one (fully connected graphs). Additionally, note that the larger the graph size $N$, the 
smaller the value of $r$ needed to approach the GOE limit. 
We would like to add that our results coincide with those in Ref.~\cite{DGK16}, where the 
$P(s)$ of RRGs with $r=0.09375$ and $r=0.3$, both with $N=1000$, was shown to be very close 
to the GOE statistics (after removing degeneracies produced by the binary nature of the adjacency 
matrices considered).

It is worth stressing that the curves shown in Fig.~\ref{Fig3}(a) have the same functional
form, but they shift to the left of the $r$-axis when the graph size is increased.
Moreover, the displacement seems constant when duplicating $N$. This observation makes
us think that these curves may obey a scaling relation (somehow dependent on $N$) with respect to the
connection radius. To look for this scaling, we first define a quantity that could characterize the displacement of the curves produced when $N$ is changed. As a matter of fact, all curves of $\beta$ vs.~$r$ in the semi-log scale look very similar: They are approximately
zero for small $r$, then grow, and finally saturate. So, we can choose as the characteristic 
quantity, for example, the value of $r$ at which the curves start to grow (i.e.~the onset of the 
delocalization transition), the value at which the curves approach the saturation (i.e.~the onset 
of the GOE limit), or the value at which the curve $d\beta/dr$ vs.~$r$ reaches its maximum 
in the transition region. We choose the last quantity, which we denote by $r^*$.
In the inset of Fig.~\ref{Fig3}(b) we present $r^*$ as a function of $N$ in a log-log scale. 
The linear trend of the data (in the log-log scale) implies a power-law relation of the form
\begin{equation}
\label{scalingEq}
r^* = \mathcal{C} N^\gamma \ .
\end{equation}
Indeed, Eq.~(\ref{scalingEq}) provides a very good fitting of the data (the fitted values are 
reported in the figure caption). Therefore, by plotting again the curves of $\beta$ now as a 
function of the connection radius divided by $N^\gamma$, as shown in Fig.~\ref{Fig3}(b), we 
observe that curves for different graph sizes $N$ collapse on top of a single curve. This means 
that once the ratio $r/N^\gamma$ is fixed, no matter the graph size, the shape of $P(s)$ is also
fixed.

Interestingly, Figure \ref{Fig3}(b) also provides a way to predict the shape of $P(s)$ of weighted RGGs 
once the ratio $r/N^\gamma$ is known:  When $r/N^\gamma<1$, $P(s)$ has the Poisson shape. 
For $r/N^\gamma>3$, $P(s)$ is practically given by the Wigner-Dyson distribution. While in the 
range of $1\le r/N^\gamma \le 3$, $P(s)$ is well described by Brody distributions characterized by 
a value of $0<\beta<1$. Thus, $r/N^\gamma\approx 1$ and $r/N^\gamma\approx 3$ mark
the onset of the delocalization transition and the onset of the GOE limit, respectively.

\subsection{Eigenfunction properties}

Once we have observed that the spectral properties of weighted RGGs show a transition from
Poisson to Wigner-Dyson statistics, we expect that the eigenfunction properties can help us to
confirm that transition.
Therefore, in order to characterize quantitatively the complexity of the 
eigenfunctions of weighted RGGs we use the Shannon entropy, which for the eigenfunction 
$\Psi^m$ is given as
\begin{equation}
\label{S}
S^m = -\sum_{n=1}^N (\Psi^m_n)^2 \ln (\Psi^m_n)^2 \ .
\end{equation}
The Shannon entropy is a measure of the number of principal components of an eigenfunction in 
a given basis. In fact, it has been already used to characterize the eigenfunctions of adjacency 
matrices of several random network models (see some examples in
Refs.~\cite{ZYYL08,GT06,JSVL10,MRP14,MARM17}).

Here, $S^m$ allows us to compute the so called entropic eigenfunction localization length~\cite{I90}, 
i.e.
\begin{equation}
\label{lH}
\ell = N \exp\left[ -\left( S_{\tbox{GOE}} - \bra S^m \ket \right)\right] \ ,
\end{equation}
where $S_{\tbox{GOE}}\approx\ln(N/2.07)$ is the entropy of a random eigenfunction with Gaussian 
distributed amplitudes. We average over all eigenfunctions of an ensemble of adjacency matrices of 
size $N$ to compute $\bra S^m \ket$ such that for each combination $(N,r)$ we use $5\times 10^5$ 
eigenfunctions. 
With definition (\ref{lH}), when $r=0$, since the eigenfunctions of the adjacency
matrices of our weighted RGGs have only one non-vanishing component with magnitude equal to one,
$\bra S^m \ket=0$ and $\ell\approx 2.07$. On the other hand, for $r=1$, 
$\bra S^m \ket=S_{\tbox{GOE}}$ and the fully chaotic eigenfunctions extend over the $N$ 
available vertices in the graph, i.e., $\ell\approx N$.

\begin{figure}[h]
\centering
 \includegraphics[scale = .4]{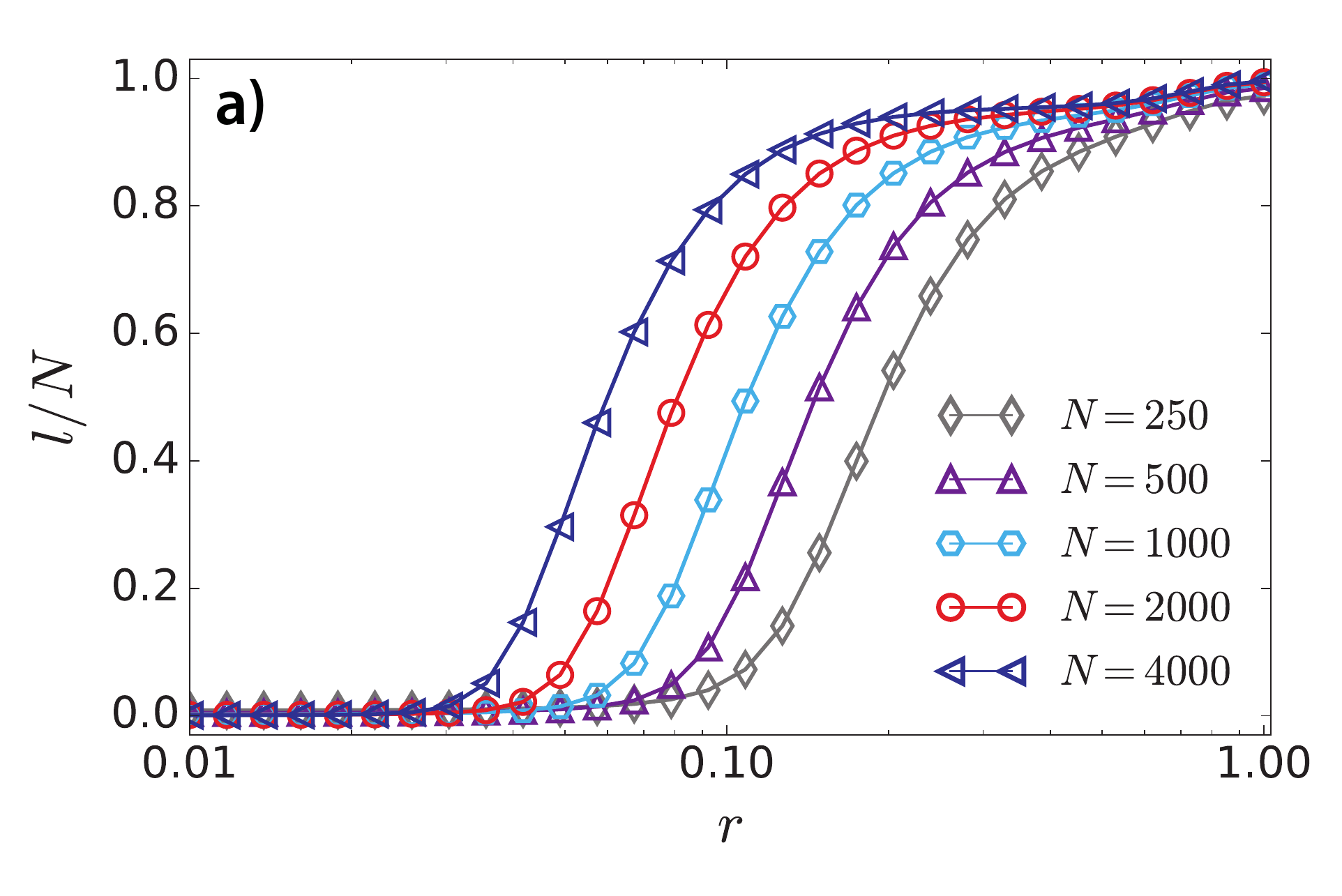}
 \includegraphics[scale = .4]{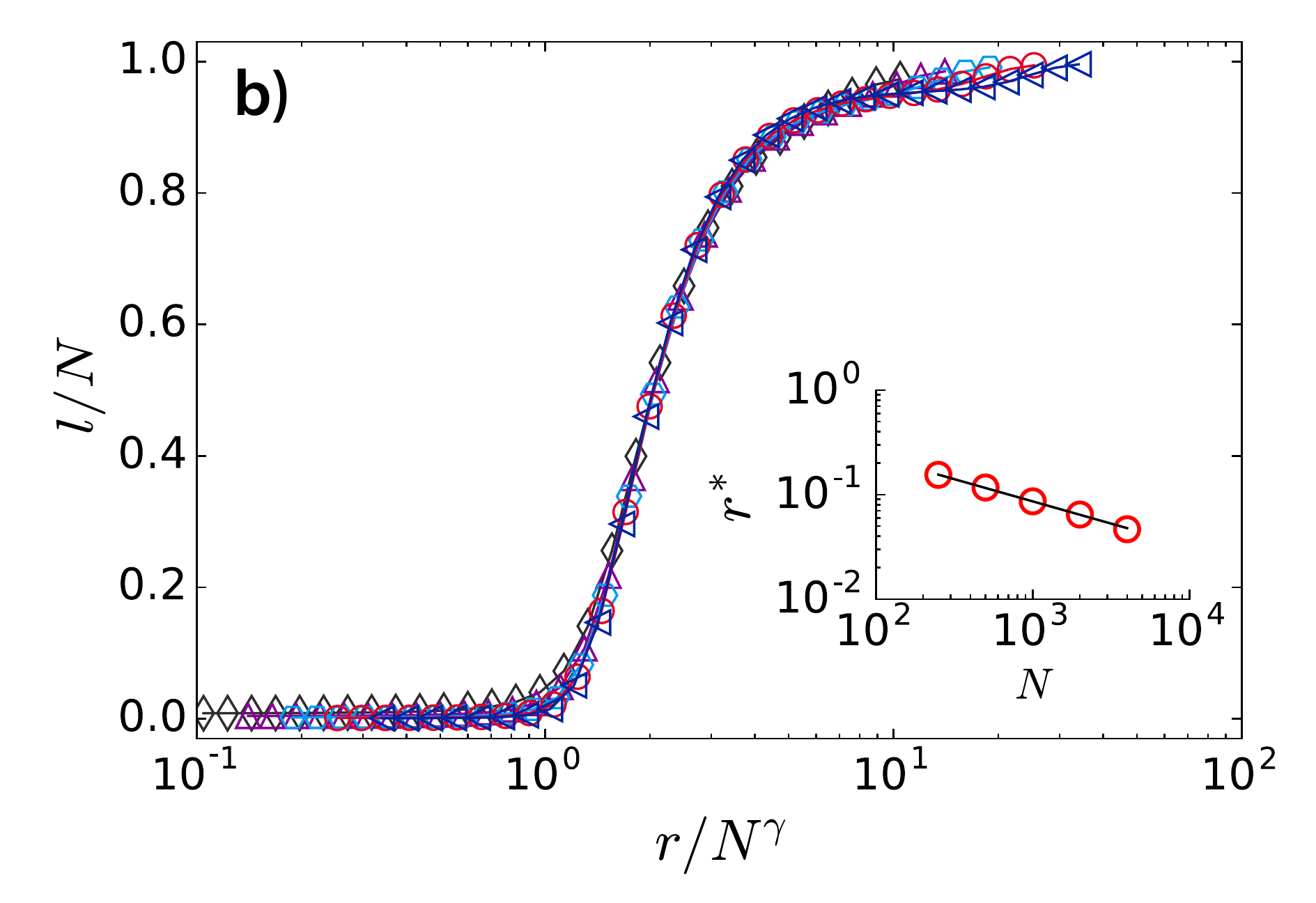}
\caption{Entropic eigenfunction localization length $\ell$ (normalized to $N$) as a function of 
(a) $r$ and (b) $r/N^\gamma$ for weighted RGGs of sizes ranging from $N=250$ to 4000. 
Inset in (b) $r^*$ vs.~$N$. Full black line is the fit of the data with Eq.~(\ref{scalingEq}) giving 
$\mathcal{C}=1.626\pm 0.061$ and $\gamma=-0.425\pm 0.006$.}
\label{Fig4}
\end{figure}

Figure~\ref{Fig4}(a) shows $\ell/N$ as a function of the connection radius $r$ for weighted RGGs of 
sizes $N=250$, 500, 1000, 2000, and 4000. We observe that the curves $\ell/N$, for different 
$N$, have the same functional form as a function of $r$ but shifted to the left for increasing $N$. 
Since this behavior is equivalent to that of $\beta$ vs.~$r$ in Fig.~\ref{Fig3}(a) we
anticipate the scaling of the curves $\ell/N$ vs.~$r/N^\gamma$. Indeed, in Fig.~\ref{Fig4}(b)
we verify the scaling by observing the coalescence of all curves onto a single one (in the
inset of Fig.~\ref{Fig4}(b) we also report the curve $r^*$ vs.~$N$ used to extract the exponent 
$\gamma$). 

From Fig.~\ref{Fig4}(b) it is clear that the curve $\ell/N$ as a function of $r/N^\gamma$ shows a 
universal behavior that can be easily described:
(i) $\ell/N$ transits from $\approx 2.07/N\sim 0$ to one by increasing $r/N^\gamma$;
(ii) for $r/N^\gamma\stackrel{<}{\sim}1$ the eigenfunctions are practically localized since $\ell\sim 1$; 
hence the delocalization transition takes place around $r/N^\gamma\approx 1$; and
(iii) for $r/N^\gamma>10$ the eigenfunctions are practically chaotic and fully extended since
$\ell\approx N$.

Note that while the delocalization transition at $r/N^\gamma\approx 1$ is observed for both
spectral and eigenfunction properties, the onset of the GOE limit occurs earlier for the $P(s)$
($r/N^\gamma\approx 3$) than for $\ell$ ($r/N^\gamma\approx 10$). This result is usual in
RMT models, see an example on random networks in Ref.~\cite{MAM15}.

\section{Weighted random rectangular graphs}

Once we have shown that spectral and eigenfunction properties of weighted RGGs (characterized 
by the Brody parameter and the entropic eigenfunction localization length, respectively) scale for the 
fixed ratio $r/N^\gamma$, we extend our study to weighted RRGs for which RGGs are a limit case.
We recall that the dimensions of the rectangular region defining the RRGs is specified by the 
parameter $a$ which we set here to 2, 4, 10, 40, 100, and 1000; such that we explore 
two-dimensional geometries, i.e., $a\sim1$, but also approach the limit of quasi-one-dimensional 
{\it wires}, when $a\gg1$.

\begin{figure*}[t]
\centering
\begin{minipage}[t]{0.48\linewidth}
\centerline{\includegraphics[scale = .4]{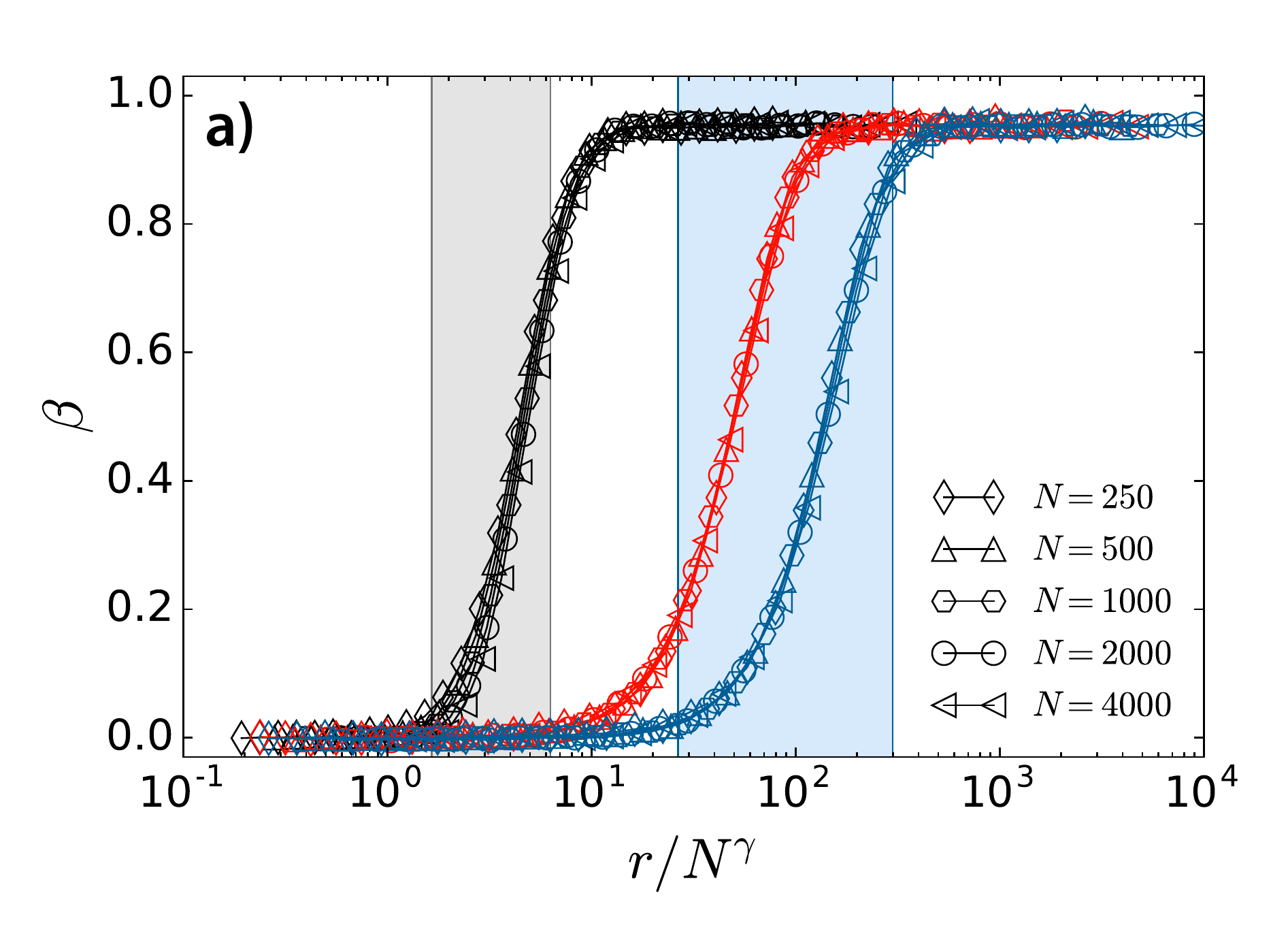}}
\centerline{\includegraphics[scale = .4]{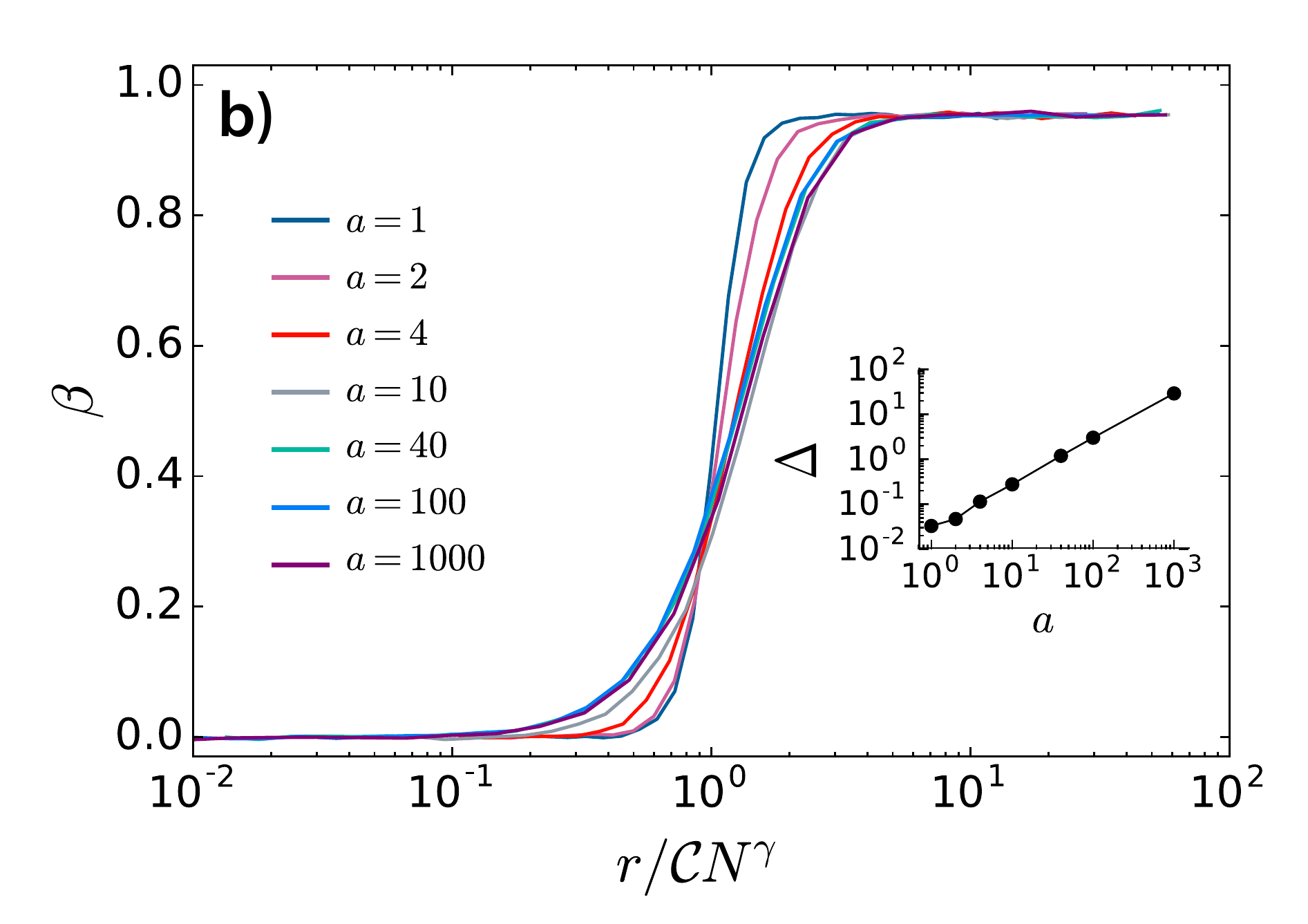}}
\centerline{\includegraphics[scale = .4]{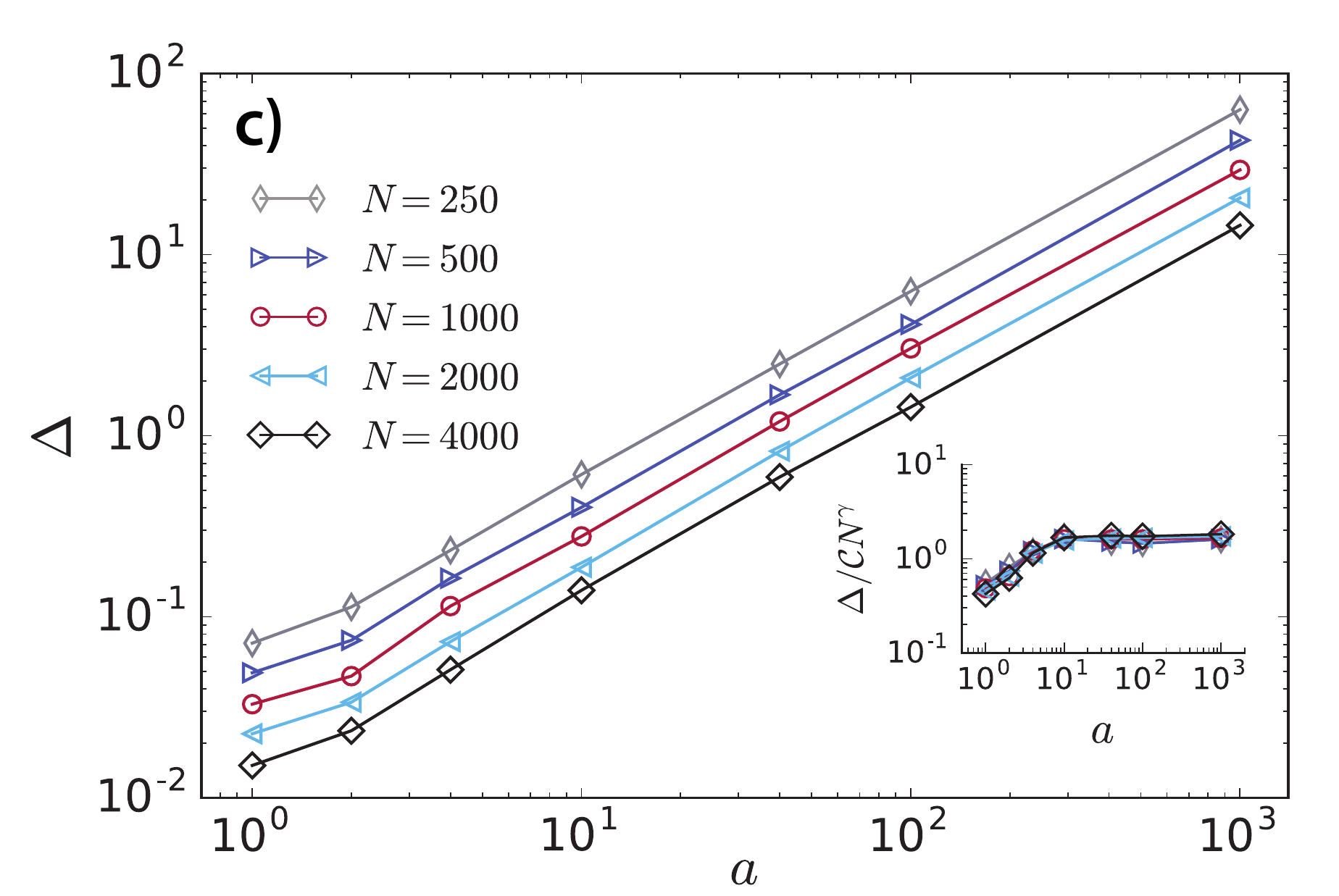}}
\caption{(Color onlie) 
(a) Brody parameter $\beta$ as a function of $r/N^\gamma$ for weighted RRGs with 
$a=4$ (left-black curves), 40 (middle-red curves), and 100 (right-blue curves).
Shaded regions depict the width of the transition region $\Delta$ (normalized to $N^\gamma$) 
for $a=4$ and 100.
(b) $\beta$ as a function of $r/{\cal C}N^\gamma$ for several values of $a$, all with $N=1000$.
Inset: $\Delta$ (from the curves in main panel) vs.~$a$.
(c) Width of the transition region $\Delta$ as a function of $a$ for several graph sizes.
Inset: $\Delta/{\cal C}N^\gamma$ vs.~$a$.
The values of ${\cal C}$ and $\gamma$ used here are reported in Table~\ref{Table1}.}
\label{Fig5}
\end{minipage}
\quad
\begin{minipage}[t]{0.48\linewidth}
\centerline{\includegraphics[scale = .4]{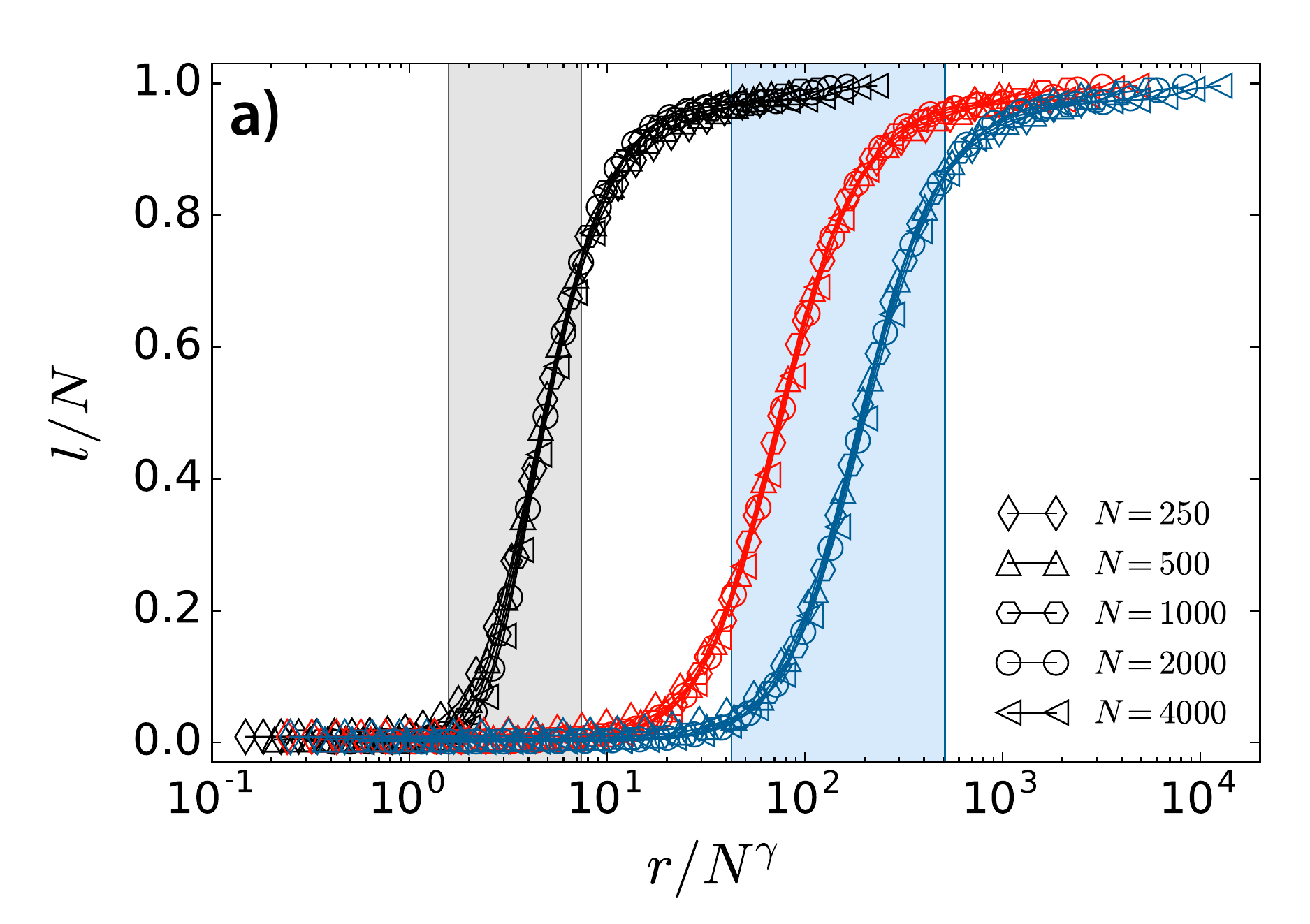}}
\centerline{\includegraphics[scale = .4]{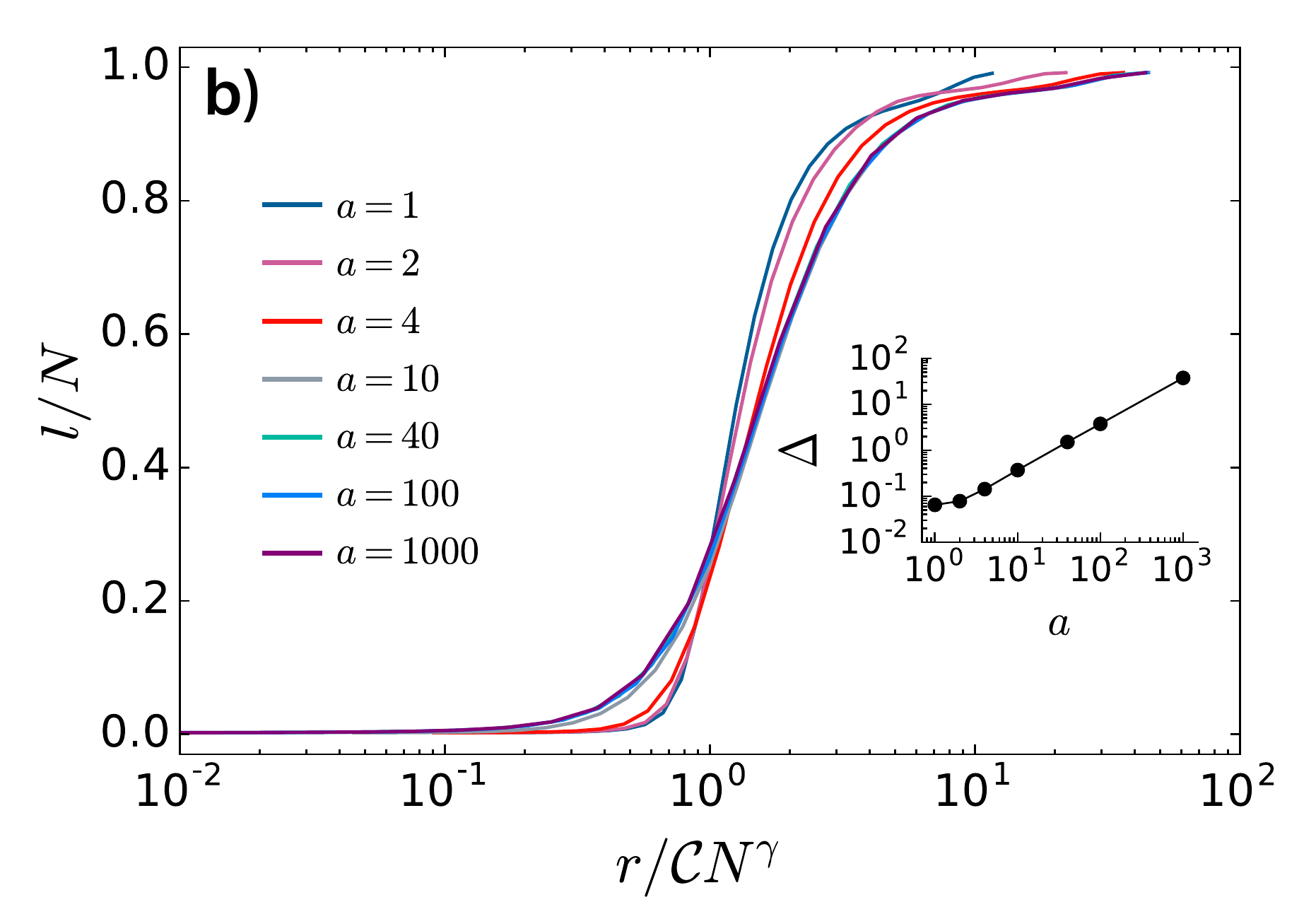}}
\centerline{\includegraphics[scale = .4]{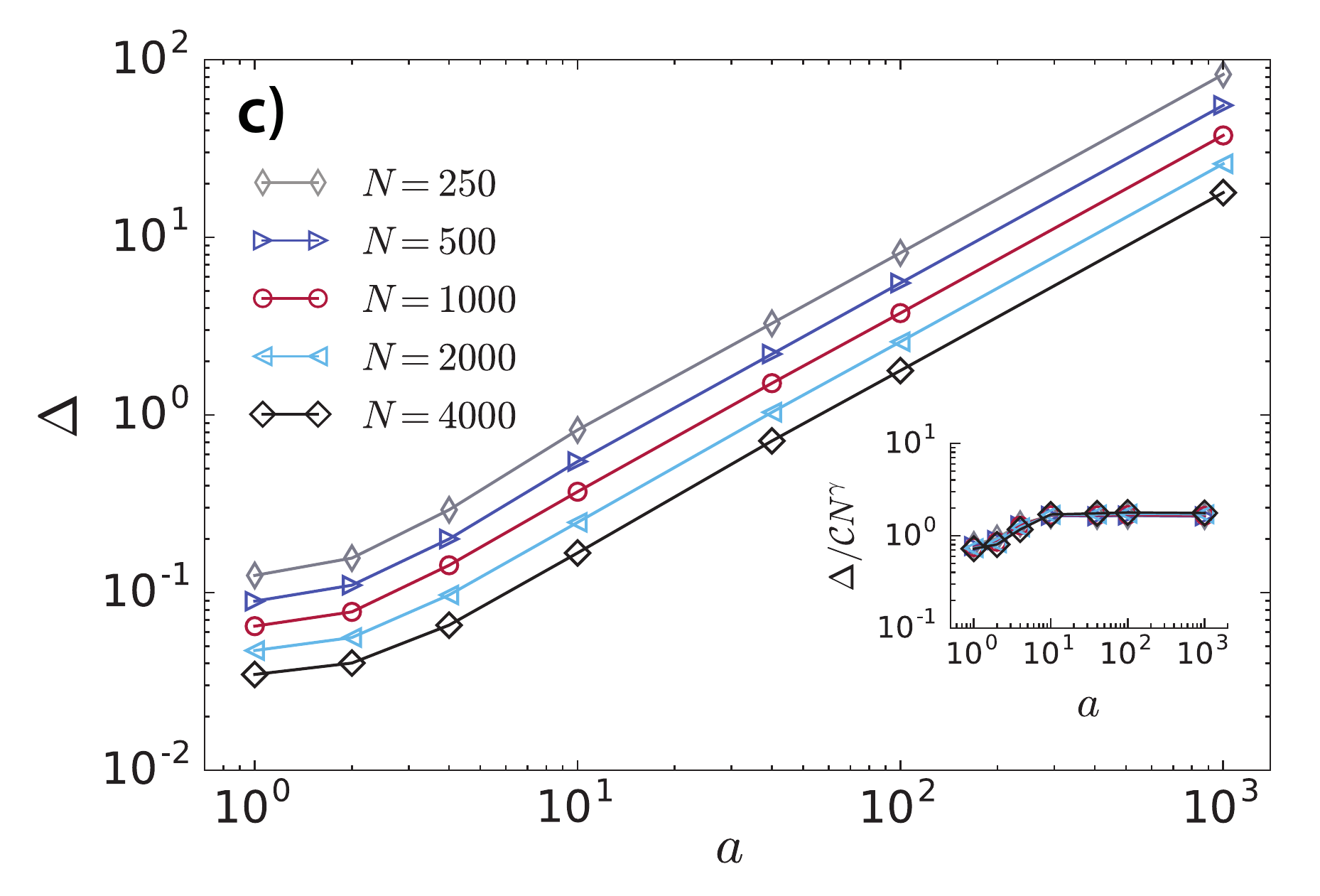}}
\caption{(Color onlie) (a) Entropic eigenfunction localization length $\ell$ (normalized to $N$)
as a function of $r/N^\gamma$ for weighted RRGs with 
$a=4$ (left-black curves), 40 (middle-red curves), and 100 (right-blue curves).
Shaded regions depict the width of the transition region $\Delta$ (normalized to $N^\gamma$) 
for $a=4$ and 100.
(b) $\ell/N$ as a function of $r/{\cal C}N^\gamma$ for several values of $a$, all with $N=1000$.
Inset: $\Delta$ (from the curves in main panel) vs.~$a$.
(c) Width of the transition region $\Delta$ as a function of $a$ for several graph sizes.
Inset: $\Delta/{\cal C}N^\gamma$ vs.~$a$.
The values of ${\cal C}$ and $\gamma$ used here are reported in Table~\ref{Table2}.}
\label{Fig6}
\end{minipage}
\end{figure*}

Based on the experience gained in the previous section on weighted RGGs we now perform a similar 
analysis for weighted RRGs: We construct several ensembles of weighted RRGs characterized by 
different combinations of $(a,r,N)$; then, after diagonalizing the corresponding adjacency matrices, 
we (i) construct histograms of $P(s)$ and, by fitting them using Eq.~(\ref{B}), extract the  
Brody parameters, and (ii) compute $\ell$ from the Shannon entropy of the eigenfunctions. 

When plotting $\beta$ and $\ell/N$ versus $r$ for any $a>1$ we observe the same scenario 
(not shown here) reported for $a=1$ in the previous Section: The curves are shifted to the left for
increasing $N$ (see Figs.~\ref{Fig3}(a) and~\ref{Fig4}(a) as a reference). Then, we foresee the scaling
of $\beta$ and $\ell$ when plotted as a function of $r/N^\gamma$. Indeed, in Figs.~\ref{Fig5}(a) 
and~\ref{Fig6}(a) we present curves of $\beta$ vs.~$r/N^\gamma$ and $\ell/N$ vs.~$r/N^\gamma$,
respectively, for several weighted RRGs with different sizes. 
In each figure we report three representative cases: $a=4$ (left-black curves), 40 (middle-red curves),
and 100 (right-blue curves). It is important to stress that we
observe a very good scaling of the curves $\beta$ and $\ell/N$ versus $r/N^\gamma$ for any $a$; 
a clear manifestation of universality. Additionally, for completeness, in Tables~\ref{Table1} and~\ref{Table2} we report the values of the
constant ${\cal C}$ and power $\gamma$ obtained from the fittings of $r^*$ vs.~$N$ with 
Eq.~(\ref{scalingEq}). 

Figures \ref{Fig5}(a) and \ref{Fig6}(a) show two important effects due to $a$:
(i) the larger the value of $a$, the larger the values of $r$ at which the onset of the delocalization 
transition and the onset of the GOE limit take place, and
(ii) the larger the value of $a$, the wider the transition region; 
for both spectral and eigenfunction properties. 
The first effect is evidenced by the displacement of the curves of $\beta$ and $\ell/N$ 
vs.~$r/N^\gamma$ to the right for increasing $a$. In fact, this displacement, quantified by
${\cal C}$, turns out to be of the order of $a$; see Tables~\ref{Table1} and~\ref{Table2}.
The second effect is better observed in Figs.~\ref{Fig5}(b) and~\ref{Fig6}(b) where we now plot 
$\beta$ and $\ell/N$ as a function of $r/{\cal C}N^\gamma$ for several values
of $a$. Here, it is clear that the transition is sharper for $a=1$ (RGGs). 
However, once $a\ge 10$ the curves normalized to ${\cal C}N^\gamma$ do not change much
and they seem to approach a limiting curve. Note that we are also including the case $a=1000$, which represents an
extremely narrow quasi-one-dimensional geometry.

We also characterize the width of the transition region (the region between the onset of the
delocalization transition and the onset of the GOE limit), that we call $\Delta$, as the full width 
at half maximum of the functions $d\beta/dr$ vs.~$r$ and $d\ell/dr$ vs.~$r$. 
As examples, the shaded regions in Figs.~\ref{Fig5}(a) and~\ref{Fig6}(a) correspond to
$\Delta$ (normalized to $N^\gamma$) for the curves with $a=4$ and 100, while in the insets of 
Figs.~\ref{Fig5}(b) and~\ref{Fig6}(b) we report the values of $\Delta$ for the curves in the 
corresponding main panels.
Finally, in Figs.~\ref{Fig5}(c) and~\ref{Fig6}(c) we plot $\Delta$ as a function of $a$ for several 
graph sizes. From these figures we observe that: 
(i) $\Delta$ increases proportional to $a$, and 
(ii) $\Delta/{\cal C}N^\gamma$ is a universal function with two regimes (see the insets):
when $1\le a < 10$, $\Delta/{\cal C}N^\gamma$ increases with $a$ while for $a\ge 10$ it 
remains constant. 

\begin{table}
  \centering
  \caption{Values of $\mathcal{C}$ and $\gamma$ obtained from the fittings of curves $r^*$ vs.~$N$
  with Eq.~(\ref{scalingEq}). Here, $r^*$ characterizes the Brody parameter as a 
  function of $r$.}
   \label{Table1}
  \begin{tabular}{|r|c|c|}
  \hline
    $a$ & $\mathcal{C}_\beta$ & $\gamma_\beta$\\
    \hline
    $1$ &     $1.678\pm 0.067$ & $-0.465\pm 0.006$ \\
    $2$ &     $1.816\pm 0.018$ & $-0.467\pm 0.002$ \\
    $4$ &     $3.781\pm 0.540$ & $-0.536\pm 0.023$ \\
    $10$ &   $6.559\pm 1.699$ & $-0.527\pm 0.042$ \\
    $40$ &   $39.12\pm 2.975$ & $-0.574\pm 0.013$ \\
    $100$ & $112.3\pm 20.51$ & $-0.592\pm 0.030$ \\
    $1000$ & $964.36\pm 18.04$ & $-0.578\pm 0.031$ \\
    \hline
  \end{tabular}
\end{table}
\begin{table}
  \centering
  \caption{Values of $\mathcal{C}$ and $\gamma$ obtained from the fittings of curves $r^*$ vs.~$N$
  with Eq.~(\ref{scalingEq}). Here, $r^*$ characterizes the entropic localization length as a 
  function of $r$.}
   \label{Table2}
  \begin{tabular}{|r|c|c|}
  \hline
    $a$ & $\mathcal{C}_\ell$ & $\gamma_\ell$\\
    \hline
    $1$ &     $1.626\pm 0.061$ & $-0.425\pm 0.006$ \\
    $2$ &     $1.811\pm 0.053$ & $-0.433\pm 0.005$ \\
    $4$ &     $3.214\pm 0.117$ & $-0.488\pm 0.006$ \\
    $10$ &   $12.61\pm 1.370$ & $-0.585\pm 0.018$ \\
    $40$ &   $48.53\pm 2.306$ & $-0.576\pm 0.008$ \\
    $100$ & $112.3\pm 20.51$ & $-0.592\pm 0.030$ \\
    $1000$ & $1282.1\pm 33.03$ & $-0.584\pm 0.004$ \\
    \hline
  \end{tabular}  
\end{table}

\section{Conclusions}

In this paper, within a random-matrix-theory approach, we have numerically studied spectral and 
eigenfunction properties of weighted RGGs and RRGs. These graph models are defined through 
the connection radius $r$, the number of vertices $N$, and, in the case of 
RRGs, the rectangle side lengths $a$ and $1/a$. Here, $0\le r \le a$.
In general, we observed a delocalization transition, in both spectral and eigenfunction properties, that 
may take place in two different ways: By increasing $r$ from zero for a fixed graph size, or
by increasing $N$ for a fixed connection radius.
Moreover, we focused on the scaling properties of $P(s)$ and $\ell$  duruing this transition, as a 
function of the graph parameters.

We first studied in detail weighted RGGs and found that:
(i) both spectral and eigenfunction properties are invariant for the ratio $r/N^\gamma$; that is,
the curves of $\beta$ and $\ell/N$ vs.~$r/N^\gamma$ fall on top of universal curves regardless of the
size of the graph;
(ii) the delocalization transition takes place at $r/N^\gamma\approx 1$; i.e.~for 
$r/N^\gamma\stackrel{<}{\sim}1$ the $P(s)$ has the Poisson shape and the eigenfunctions are 
practically localized since $\ell\sim 1$;
(iii) the onset of the GOE limit takes place at $r/N^\gamma\approx 3$ for the spectral properties
and at $r/N^\gamma\approx 10$ for the eigenfunction properties; this means that $P(s)$ is very 
close to the Wigner-Dyson distribution when $r/N^\gamma > 3$, while for $r/N^\gamma>10$ 
the eigenfunctions are fully extended since $\ell\approx N$. Here, $\gamma=-0.465\pm 0.006$ 
for spectral and $\gamma=-0.425\pm 0.006$ for eigenfunction properties.

Then, we analyzed weighted RRGs where, in fact, the case of weighted RGGs is obtained for $a=1$. It is worth remarking that although we considered RRGs embedded in two-dimensional geometries, i.e.~$a\sim 1$, we also approached the limit of quasi-one-dimensional wires with $a=1000\gg1$. Here, we showed that 
(i) both spectral and eigenfunction properties are invariant for the ratio $r/N^\gamma$ for any $a$;
(ii)  when increasing $a$, the values of $r$ at which the onsets of the delocalization 
transition and of the GOE limit take place also increase; and
(iii) the width of the transition region $\Delta$ between the onset of the delocalization transition and that of the GOE limit varies as 
$\Delta\propto a$.

In addition we were able to identify two regimes for RRGs that we call the two-dimensional (2D) 
regime, when $1\le a<10$, and the quasi-one-dimensional (Q1D) regime, for $a\ge 10$.
In the Q1D regime, we have demonstrated that the spectral and eigenfunction properties are 
universal for the fixed ratio $r/{\cal C}N^\gamma$, with ${\cal C}\sim a$; that is,
the curves of $\beta$ and $\ell/N$ vs.~$r/{\cal C}N^\gamma$ are invariant. In this regime,
the exponent $\gamma$ does not depend on $a$ anymore and takes the value 
$\gamma\approx -0.58$. In this sense, the 2D regime showed a richer behavior since in this case quantities such as $\gamma$ and $\Delta$ (even normalized to ${\cal C}N^\gamma$) do change with $a$.

Overall, our results might shed additional light on the critical properties and structural organization
of spatially embedded systems that can be mapped into networks. For instance, our findings may provide hints to design systems with desired localization properties and to better understand critical properties that depend on eigenfunction properties. This could be done by tuning the parameters of the random rectangular graph. It would also be of interest to explore these issues in adaptive spatial networks, for example, by coupling a random walk model that allows nodes to move in and out of the connection radius of their neighbors, thus dynamically tuning in an effective way the density of the nodes within $r$ and eventually the regime at which the system operates regarding its spectral and eigenfunction properties. We hope that our work inspire such studies in the near future.

\begin{acknowledgments}
This work was partially supported by 
VIEP-BUAP (Grant No.~MEBJ-EXC17-I), 
Fondo Institucional PIFCA (Grant No.~BUAP-CA-169), and 
CONACyT (Grant No.~CB-2013/220624). Y. M. acknowledges partial support from the Government of Arag\'on, Spain through a grant to the group FENOL, and by MINECO and FEDER funds (grant FIS2014-55867-P).
\end{acknowledgments}


\begin{thebibliography}{99}

\bibitem{review2006} 
S. Boccaletti, V. Latora, Y. Moreno, M. Chavez, and D. U. Hwang, 
Complex networks: Structure and dynamics,
Phys. Rep. {\bf 424}, 175 (2006).

\bibitem{B11} 
M. Barth\'el\'emy, 
Spatial networks, 
Phys. Rep. {\bf 499}, 1 (2011).

\bibitem{DC02} 
J. Dall and M. Christensen, 
Random geometric graphs, 
Phys. Rev. E {\bf 66}, 016121 (2002).

\bibitem{P03} 
M. Penrose, 
{\it Random Geometric Graphs} 
(Oxford University Press, Oxford, 2003).

\bibitem{G59} 
E. N. Gilbert, 
Random graphs, 
Ann. Math. Stat. {\bf 30}, 1141 (1959).

\bibitem{WG09} 
P. Wang and M. C. Gonzalez, 
Understanding spatial connectivity of individuals with non-uniform population density,
Philos. Trans. R. Soc., A {\bf 367}, 3321 (2009).

\bibitem{DGMN09} 
A. Diaz-Guilera, J. Gomez-Gardenes, Y. Moreno, and M. Nekovee, 
Synchronization in random geometric graphs,
Int. J. Bifurc. Chaos {\bf 19}, 687 (2009).

\bibitem{Nekovee07} 
M. Nekovee, 
Worm epidemics in wireless ad hoc networks,
New J. Phys. {\bf 9}, 189 (2007).

\bibitem{TG07} 
Z. Toroczkai and H. Guclu, 
Proximity networks and epidemics,
Physica A {\bf 378}, 68 (2007).

\bibitem{BAK08} 
C. P. Brooks, J. Antonovics, and T. H. Keitt, 
Spatial and temporal heterogeneity explain disease dynamics in a spatially explicit network model,
Am. Nat. {\bf 172}, 149 (2008).


\bibitem{ES15}
E. Estrada and M. Sheerin,
Random rectangular graphs,
Phys. Rev. E {\bf 91}, 042805 (2015).

\bibitem{EC15}
E. Estrada and G. Chen, 
Synchronizability of random rectangular graphs,
CHAOS {\bf 25}, 083107 (2015).

\bibitem{ES16}
E. Estrada, M. Sheerin, 
Consensus dynamics on random rectangular graphs,
Physica D {\bf 323-324}, 20 (2016).

\bibitem{EMSM16}
E. Estrada, S. Meloni, M. Sheerin, and Y. Moreno,
Epidemic spreading in random rectangular networks,
Phys. Rev. E {\bf 94}, 052316 (2016).

\bibitem{BEJ07}
P. Blackwell, M. Edmondson-Jones, and J. Jordan,
Spectra of adjacency matrices of random geometric graphs,
Research report: Dept. of Probability \& Statistics, University of Sheffield, no. 570/07, 2007.
(http://www.jonathanjordan.staff.shef.ac.uk/rgg.html).

\bibitem{NGB14}
A. Nyberg, T. Gross, K. E. Bassler,
Mesoscopic structures and the Laplacian spectra of random geometric graphs,
J Complex Netw {\bf 3}(4), 543-551 (2015).

\bibitem{DGK16}
C. P. Dettmann, O. Georgiou, and G. Knight,
Spectral statistics of random geometric graphs,
Europhys. Lett. {\bf 118}, 18003 (2017).

\bibitem{BBPV04}
A. Barrat, M. Barth\'elemy, R. Pastor-Satorras, and A. Vespignani,
The architecture of complex weighted networks,
PNAS {\bf 101}, 3747 (2004).



\bibitem{metha}
M. L. Metha,
{\it Random matrices} (Elsevier, Amsterdam, 2004).

\bibitem{EE92}
S. N. Evangelou and E. N. Economou,
Spectral density singularities, level statistics, and localization in a sparse random matrix ensemble,
Phys. Rev. Lett. {\bf 68}, 361 (1992).

\bibitem{JMR01}
A. D. Jackson, C. Mejia-Monasterio, T. Rupp, M. Saltzer, and T. Wilke,
Spectral ergodicity and normal modes in ensembles of sparse matrices,
Nucl. Phys. A {\bf 687}, 405 (2001).

\bibitem{ZX00}
C. P. Zhu and S. J. Xiong,
Localization-delocalization transition of electron states in a disordered quantum small-world network,
Phys. Rev. B {\bf 62}, 14780 (2000).

\bibitem{GGS05}
O. Giraud, B. Georgeot, and D. L. Shepelyansky,
Quantum computing of delocalization in small-world networks,
Phys. Rev. E {\bf 72}, 036203 (2005).

\bibitem{SKHB05}
M. Sade, T. Kalisky, S. Havlin, and R. Berkovits,
Localization transition on complex networks via spectral statistics,
Phys. Rev. E {\bf 72}, 066123 (2005).

\bibitem{JKBH08}
L. Jahnke, J. W. Kantelhardt, R. Berkovits, and S. Havlin,
Wave localization in complex networks with high clustering,
Phys. Rev. Lett. {\bf 101}, 175702 (2008).

\bibitem{BJ07}
J. N. Bandyopadhyay and S. Jalan,
Universality in complex networks: Random matrix analysis,
Phys. Rev. E {\bf 76}, 026109 (2007).

\bibitem{JB08}
S. Jalan and J. N. Bandyopadhyay,
Random matrix analysis of network Laplacians,
Physica A {\bf 387}, 667 (2008).

\bibitem{ZYYL08}
G. Zhu, H. Yang, C. Yin, and B. Li,
Localizations on complex networks,
Phys. Rev. E {\bf 77}, 066113 (2008).

\bibitem{J09}
S. Jalan,
Spectral analysis of deformed random networks
Phys. Rev. E {\bf 80}, 046101 (2009).

\bibitem{MAM15}
J. A. Mendez-Bermudez, A. Alcazar-Lopez, A. J. Martinez-Mendoza, 
F. A. Rodrigues, and T. K. DM. Peron,
Universality in the spectral and eigenfunction properties of random networks,
Phys. Rev. E {\bf 91}, 032122 (2015).

\bibitem{B73}
T. A. Brody,
A statistical measure for the repulsion of energy levels,
Lett. Nuovo Cimento {\bf 7}, 482 (1973).

\bibitem{B81}
T. A. Brody, J. Flores, J. B. French, P. A. Mello, A. Pandey, and S. S. M. Wong,
Random-matrix physics: spectrum and strength fluctuations,
Rev. Mod. Phys. {\bf 53}, 385 (1981).

\bibitem{JB07}
S. Jalan and J. N. Bandyopadhyay,
Random matrix analysis of complex networks,
Phys. Rev. E {\bf 76}, 046107 (2007).

\bibitem{GT06}
L. Gong and P. Tong,
von Neumann entropy and localization-delocalization transition of electron 
states in quantum small-world networks,
Phys. Rev. E {\bf 74}, 056103 (2006).

\bibitem{JSVL10}
S. Jalan, N. Solymosi, G. Vattay, and B. Li,
Random matrix analysis of localization properties of gene coexpression network,
Phys. Rev. E {\bf 81}, 046118 (2010).

\bibitem{MRP14}
G. Menichetti, D. Remondini, P. Panzarasa, R. J. Mondragon, and G. Bianconi,
Weighted multiplex networks,
PLoS ONE {\bf 9}, e97857 (2014).

\bibitem{MARM17}
J. A. Mendez-Bermudez, G. F. de Arruda, F. A. Rodrigues, and Y. Moreno,
Scaling properties of multilayer random networks,
Phys. Rev. E {\bf 96}, 012307 (2017).

\bibitem{I90}
F. M. Izrailev,
Simple models of quantum chaos: Spectrum and eigenfunctions,
Phys. Rep. {\bf 196}, 299 (1990).

\end{thebibliography}
\end{document}